\documentclass{article}
\usepackage{arxiv}

\usepackage[utf8]{inputenc} 
\usepackage[T1]{fontenc}    
\usepackage{hyperref}       
\usepackage{url}            
\usepackage{booktabs}       
\usepackage{amsfonts}       
\usepackage{nicefrac}       
\usepackage{microtype}      
\usepackage{lipsum}
\usepackage{graphicx}

\title{Representation, Exploration and Recommendation of Music Playlists}

\author{
  Piyush Papreja\hspace{5mm} Hemanth Venkateswara\hspace{5mm} Sethuraman Panchanathan \\
  Center for Cognitive Ubiquitous Computing (CUbiC)\\
  Arizona State University, Tempe, AZ 85281\\
  \texttt{\{ppapreja, hemanthv, panch\}@asu.edu}
 } 
 
\begin{document}
\maketitle

\begin{abstract}
Playlists have become a significant part of our listening experience because of the digital cloud-based services such as Spotify, Pandora, Apple Music. Owing to the meteoric rise in the usage of playlists, recommending playlists is crucial to music services today. Although there has been a lot of work done in playlist prediction, the area of playlist representation hasn't received that level of attention. Over the last few years, sequence-to-sequence models, especially in the field of natural language processing, have shown the effectiveness of learned embeddings in capturing the semantic characteristics of sequences. We can apply similar concepts to music to learn fixed length representations for playlists and use those representations for downstream tasks such as playlist discovery, browsing, and recommendation. In this work, we formulate the problem of learning a fixed-length playlist representation in an unsupervised manner, using Sequence-to-sequence (Seq2seq) models, interpreting playlists as sentences and songs as words. We compare our model with two other encoding architectures for baseline comparison. We evaluate our work using the suite of tasks commonly used for assessing sentence embeddings, along with a few additional tasks pertaining to music, and a recommendation task to study the traits captured by the playlist embeddings and their effectiveness for the purpose of music recommendation.
\end{abstract}


\section{Introduction}
In this age of cloud-based music streaming services such as Spotify\cite{spotify}, Pandora\cite{pandora}, Apple music\cite{apple-music} among others, with millions of songs at their fingertips, users have grown accustomed to, a) immediate attainment of their music demands, and b) an extended experience \cite{choi2016towards}. Recommendation engines service one aspect of this change in user behavior. They help users find new music based on their preferences. Playlists handle the second aspect of the changing behavior, which is the need for an extended experience. An extended experience is achieved by sustaining the mood of the songs in a playlist. For example, Spotify has over two billion playlists \cite{billionplaylist} created for every kind of mood (sad, happy, angry, calm, etc.), activity (running, workout, studying, etc.), and genre (blues, rock, pop, etc.).

Over the past couple of years, the playlist recommendation task has become analogous to playlist prediction/creation and continuation rather than playlist discovery. However, playlist discovery forms a significant part of the overall playlist recommendation pipeline, as it is an effective way to help users discover existing playlists on the platform by leveraging nearest-neighbor techniques. The aim of this work is to create an end-to-end pipeline for learning playlist embeddings which can be directly used for recommendation purposes. In the past few years, sequence-to-sequence learning \cite{sutskever2014sequence} has been widely used to learn effective sentence embeddings in applications like neural machine translation \cite{bahdanau2014neural}. We make use of the relationship playlist:songs :: sentences:words, and take inspiration from research in natural language processing to model playlist embeddings the way sentences are embedded. 

\subsection{Why playlist embeddings?}
Current research pertaining to playlist is in the areas of automatic playlist generation \cite{andric2006automatic}, \cite{logan2002content} \cite{chen2012playlist}, and continuation \cite{chen2018recsys} \cite{volkovs2018two} \cite{yang2018mmcf}. Multiple solutions have been proposed to address these problems, like reinforcement learning \cite{liebman2015dj} and Recurrent Neural Network-based models \cite{choi2016towards} for playlist generation and playlist continuation tasks. However, great success has been achieved in the field of natural language processing using the power of learned embeddings. These fixed-length embeddings are easier to use and manipulate and can be used for tasks such as machine translation and query-and-search. A case can be made for using similar methods for music playlists, as the semantic properties captured can be leveraged for providing with good-quality recommendations. It can be easily integrated with other modes of information such as word2vec \cite{mikolov2013efficient} model, or content-analysis-based models \cite{lee2009unsupervised} or a combination of both, thus providing a multi-modal recommendation. Another use case for projecting the playlists onto an embedding space is easier browsing (like MusicBox\cite{lillie2008musicbox} does for songs) through the entire corpus. It can be used to discover playlists that do not exactly fit into one genre and hence are difficult to find for using the conventional query-and-search method.
Lastly, variational sequence-to-sequence models \cite{zhang2016variational}, along with added contextual information can be used for generating playlists from the embedding space \cite{bowman2015generating}.

\subsection{Why unsupervised learning?}
One of the major challenges when working with playlists is the lack of labeled data. Natural language processing has many popular supervised datasets such as SNLI \cite{bowman2015large}, Microsoft's paraphrase detection \cite{dolan2004unsupervised}, SICK dataset \cite{marelli2014semeval}, that are used to learn sentence embeddings that capture distinct discriminative characteristics which would not be possible without the labeled data. In the absence of annotated playlist datasets we resort to unsupervised learning to model playlist representations.

\subsection{Our Contribution}
In this work, we learn playlist embeddings using unsupervised learning and perform a comprehensive analysis of the embeddings learned by different encoders. We consider two kinds of models for this work: a) Seq2seq models and b) Bag of Words (BoW) models. We compare different embedding sizes, and architectures for each of the models, exploring their effect on the quality of resulting representation. We evaluate the embedding models by testing for the extent of genre information captured in the playlist embedding, along with other standard sentence-based characteristics such as length, order, content information. We also evaluate the effectiveness of the proposed embedding models using a recommendation task. To the best of our knowledge, our work is the first attempt at modeling and extensively analyzing compact playlist representations for playlist recommendation purposes, with inspirations from natural language processing. 

\section{Related Work}
With regards to this work, natural language processing and music have important similarities. Both have sequential structure in their constituent parts - words in a sentence are akin to audio segments in a song or songs in a playlist. Both have semantic relationships between the elements of the sequence. Due to these similarities, there have been many works which employ techniques from the field of natural language processing by translating the problem to an already solved-problem in natural language processing, like \cite{mcfee2011natural}, which uses this analogy for evaluating automatically generated playlists. \\
Embedding models are often used alongside the aforementioned approach in Music Information Retrieval (MIR) to project the data onto a compact space. Zheleva et al \cite{zheleva2010statistical} create a statistical model for capturing user taste using Latent Dirichlet Allocation (LDA) \cite{blei2003latent}. Collaborative Filtering (CF) \cite{herlocker1999algorithmic} is a widely used method for recommendation, but it suffers from shortcomings such as lack of consideration for order of items in the list and that there is no way to adjust the search results based on query \cite{chen2016query}. Similar to CF, there have been many neural network based works \cite{chen2016query} \cite{van2013deep} \cite{lee2018deep}  which project the corpus and the user profiles to a low dimensionality vector and then recommend items based on cosine similarity between the query and the corpus items. But for this work we are concerned with works which do not have users in the loop. Volkovs et al \cite{volkovs2018two} create a playlist embedding, albeit with a task-tailored objective function for automatic playlist continuation. Yang et al \cite{yang2018mmcf} use a custom autoencoder with an aim to make it easier to include multiple modalities in the input. Ludewig et al \cite{ludewig2018effective} use tf-idf \cite{ramos2003using} to create playlist embeddings. Aalto \cite{aalto2015learning} uses eigenvectors from the playlist to create a representation and further uses cosine similarity between the playlists to compare playlists. However, one of the major limitations of this approach is that it doesn't take into account the order of items in the playlist.  \\
We also briefly look at the work done with regards to seq2seq learning since we apply this technique to our work. Cho et al \cite{cho2014learning} used an RNN-based network to create fixed length representation of variable length source and target sequences. However, RNNs are not good enough for capturing long-term dependencies. Sutskever et al \cite{sutskever2014sequence} improved on this by using LSTM units instead of RNN, and using separate networks for encoder and decoder to increase the model capacity. Further improvement was made on this in \cite{bahdanau2014neural} by enabling the model to translate even longer sequences by introduction of Attention mechanism, which is a technique used by the decoder to make use of source sequence in making the output prediction. 
\\ Our work is different from the previous works in sense that our aim is to leverage the playlist embeddings and use those for recommending existing playlists, hence making playlists as the prime recommendation items. In addition to that, we also focus on creating a framework of evaluation tasks for the playlist embeddings, which being source agnostic, can be used to evaluate other models in the future as well.

\section{Sequence to Sequence Learning} \label{seq2seq-primer}
Here we briefly describe the RNN Encoder–Decoder framework, proposed first in  \cite{cho2014learning} and later improved in \cite{sutskever2014sequence}, upon which our model is based. Given a sequence of input vectors $x = \{x\textsubscript{1}, x\textsubscript{2}, x\textsubscript{3}...x\textsubscript{T}\}$, the encoder reads this sequence and outputs a vector $c$ called the context vector. The context vector represents a compressed version of the input sequence which is then fed to the decoder which predicts tokens from the target sequence. One of the significant limitations of this approach was that the model was not able to capture long term dependencies for relatively longer sequences  \cite{bengio1994learning}. This problem was partially mitigated in \cite{sutskever2014sequence} by using LSTM \cite{sundermeyer2012lstm} units instead of vanilla RNN units and feeding the input sequence in the reversed order to solve for lack of long-term dependency capture. \\
Bahdanau et al \cite{bahdanau2014neural} introduced the attention mechanism to solve this problem which involved focusing on a specific portion of the input sequence when predicting the output at a particular time step. The attention mechanism ensures the encoder doesn't have to encode all the information into a single context vector. In this setting, the context vector $c$ is calculated using weighted sum of hidden states $h\textsubscript{j}$:

\begin{equation}
    c_{i}=\sum_{j=1}^{T_{x}} \alpha_{i j} h_{j}
\end{equation}

where $\alpha_{i j}$ is calculated as follows:

\begin{equation}
    \alpha_{i j}=\frac{\exp \left(e_{i j}\right)}{\sum_{k=1}^{T_{x}} \exp \left(e_{i k}\right)}
\end{equation}

where $e_{i j}=a\left(s_{i-1}, h_{j}\right)$ and $s\textsubscript{i-1}$ is the decoder state at time step $i-1$ and $h_j$ is the encoder state at time step $j$. $a(.)$ is the alignment model which scores how well the output at time step $i$ aligns with the input at time step $j$. The alignment model $a$ is a shallow feed forward neural network which is trained along with the rest of the network.

\section{Embedding Models}
We consider variants of two main kinds of models for learning playlist representations, which we present in this section: 

\subsection{Bag-of-Words Models (BoW):}
BoW models create a representation of the sequence by simply aggregating the constituent items representations. They don't take into account the order of the items in the sequence. We use these models for baseline comparison.

\begin{enumerate}
        \item  \textbf{Base Bag-of-Words Model (base-BoW):} Given a sentence $s$ having a collection of words $\{w_1, w_2, \ldots, w_m\}$, the sentence embedding is calculated using a simple arithmetic mean of its constituent word embeddings. The effectiveness of this approach coupled with the simplicity of computation makes it a very competitive baseline for comparison.
    
      \item \textbf{Weighted Bag-of-Words Model (weighted-BoW):} Introduced in \cite{arora2016simple}, it uses a weighted averaging scheme to get the sentence embedding vectors followed by their modification using singular-value decomposition (SVD). This method of generating sentence embeddings proves to be a stronger baseline compared to the traditional averaging.
\end{enumerate}

\subsection{Seq2seq Models (Seq2seq):}
These models are based on the the RNN-Encoder-Decoder framework discussed in Section \ref{seq2seq-primer}.
  Given a sequence of $n$ words $x_{t=1,\ldots,n}$, an RNN-encoder generates hidden states $h_{t=1,\ldots,n}$. The attention mechanism takes in these hidden states along with the decoder states and outputs context vectors {$c\textsubscript{t=1,...,n}$}, which are then fed to the decoder to predict the output {$y\textsubscript{t=1,...,n}$}. 
  
  \begin{enumerate}
      \item \textbf{Base Unidirectional Encoder (base-Seq2seq):} We use a unidirectional RNN, and global attention model \cite{luong2015effective} as our base seq2seq model.
      
       \item \textbf{Bidirectional Encoder (bi-Seq2seq):} For this model, the encoder generated hidden state h\textsubscript{t} where t $\in\{1,..n\}$ is the concatenation of a forward RNN and a backward RNN that read the sentences in two opposite directions. We use global attention for this one as well.
  \end{enumerate}

\section{Experimental Setup}

\subsection{Corpus}

\subsubsection{Data Source}
We created the corpus by downloading publicly available playlists from Spotify using the Spotify developer API \cite{spotifydev}. We downloaded 1 million playlists\footnote{Data statistics details are mentioned in the supplementary material.} from Spotify, consisting of both user-created playlists as well as Spotify-curated ones. 

\subsubsection{Data Filtering}
As a part of cleaning up the data before training, we follow \cite{de2018large} in discarding the less frequent, and less relevant items from our dataset. Tracks occurring in less than 3 playlists are discarded. Playlists with the less than 30\% of tracks left after this are also removed. All duplicate tracks from playlists are removed. And finally, only playlists with lengths in the range [10-5000] are retained and the rest are discarded. This leaves us with a total of 745,543 unique playlists and 2,470,756 unique tracks and 2680 unique genres.

\subsubsection{Data Labeling: Genre Assignment}

The songs in our dataset do not have genre labels, however artists do. Despite there being a 1:1 mapping between artist and song, we do not use the artist genres for songs as it is because:

\begin{itemize}
    \item  An artist can have songs of different genres.
    \item  Since genres are subjective in their nature (\textit{rock v/s soft-rock v/s classic rock}), having such high number of genres for songs would result in an ambiguity between the genres with respect to empirical evaluation (classification) and add to the complexity of the problem.
\end{itemize}{}

Hence, we aim to bring down the number of genres such that they are relatively mutually disjoint (such as \textit{rock, electronic, classical} etc.) \\

To achieve this, we train a word-2-vec model\cite{mikolov2013efficient}\footnote{For details about the word2vec algorithm configuration, please refer to the supplementary material.} on our corpus to get song embeddings, which capture the semantic characteristics (such as genre) of the songs by virtue of their co-occurrence in the corpus. Separate models are trained for embedding sizes k $\in\{500,750,1000\}$. For each of the embedding sizes, the resulting song embeddings are then clustered into 200\footnote{This number was chosen with an aim to get the maximum feasible clusters (and minimum songs per cluster) while keeping the number within a limit which was feasible for annotating the data.} clusters. For each cluster:
\begin{itemize}
    \item Artist genre is applied to each corresponding song and a genre-frequency (count) dictionary is created. A sample genre-count dictionary for cluster with 17 songs would look like  \{rock: 5, indie-rock:3, blues: 2, soft-rock: 7\}
    
    \item From this dictionary, the genre having a clear majority\footnote{This was a subjective decision. For example, a dictionary having \{rock: 5, indie-rock:3, blues: 2, soft-rock: 7\} is assigned the genre \textit{rock}} is assigned as the genre for all the songs in that cluster.
    
    \item All the songs in a cluster with no clear genre majority are discarded for annotation.
\end{itemize}{}

\begin{figure}
 \centerline{\framebox{
 \includegraphics[scale=0.25]{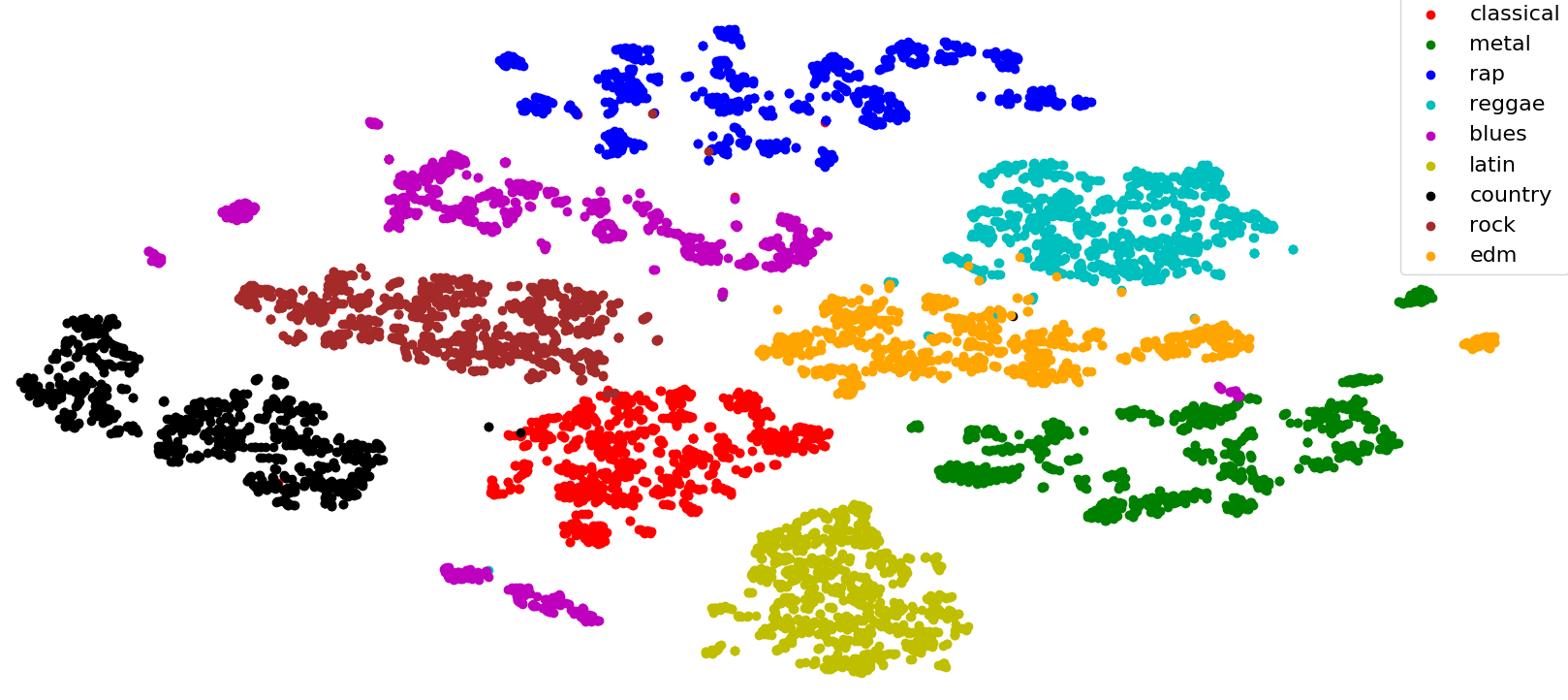}}}
 \caption{t-SNE plot for genre-annotated songs for embedding size 750, with 1000 sampled songs for each genre}
 \label{fig:tsne-songs}
\end{figure}

 Based on the observed genre-distribution in the data, and as a result of clustering sub-genres (such as \textit{soft-rock}) into parent genres (such as \textit{rock}), the genres finally chosen for annotating the clusters are: \textit{Rock, Metal, Blues, Country, Classical, Electronic, Hip Hop, Reggae and Latin}. To validate our approach, we train a classifier on our dataset consisting of annotated song embeddings. With training and test set kept separate at the time of training, we achieve a 94\% test\footnote{Result achieved for embedding size 750. Comparable results achieved for other sizes as well.} accuracy. We also generate a t-SNE plot of the annotated songs to further validate our approach as shown in Figure \ref{fig:tsne-songs}. \\
For \textbf{playlist-genre annotation}, only the playlists having all the songs annotated, are considered for annotation. This leaves us with 339,998 playlists in total. Further, only those playlists are assigned genres for which more than 70\% of the songs agree on a genre resulting in 35,527 genre-annotated playlists.

\subsection{Training: Configuration }
\begin{enumerate}
   \item \textbf{weighted-BoW Model:} For the base configuration of the model, we set the value of the parameter $a$ to $e^{-3}$ where the weight given to each word in the corpus is $a /(a+p(w))$ and $a$ is the controlled parameter.
  We experiment with different values for $a$ ranging from $e^{-3}$ and $e^{-5}$.
  
  \item \textbf{Seq2seq Models:} All of our Seq2seq encoders use a 3 layer network with hidden state size controlled for and k $\in\{500,750,1000\}$ . We experimented with both LSTM and GRU units. We tried Adam and SGD\footnote{SGD performed generally worse than Adam, hence the details for SGD are not included} optimizers. We also set the maximum gradient norm to 1 to prevent exploding gradients.
\end{enumerate}

\subsection{Training: Task}  
We train seq2seq models on our dataset as autoencoders where the target sequence is the same as the source sequence (in our case, a playlist), and the goal of the model is to reconstruct the input sequence. For more training details, please refer to the supplementary material.

\begin{figure*}
 \centerline{\framebox{
 \includegraphics[width=\textwidth]{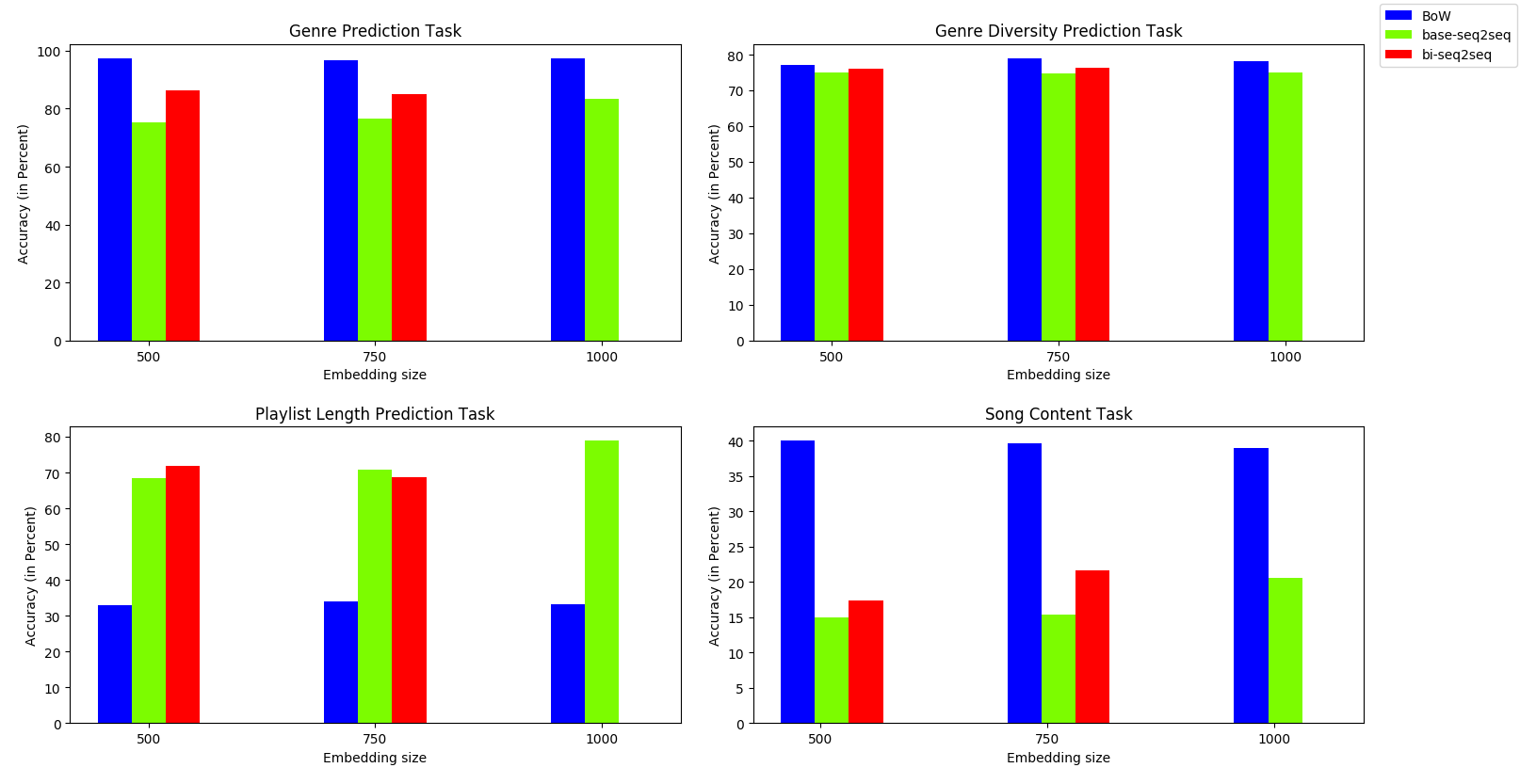}}}
 \caption{Evaluation (embedding probing) task results with respect to the encoder hidden state size. Missing bars are for cases where it was not possible to train an embedding of that size due to memory constraints.}
 \label{fig:accuracy-emb-size}
\end{figure*}

\section{Evaluation Tasks}
Ideally, a good playlist embedding should encode information about the genre of the songs it contains, the order of songs, length of playlist and songs themselves. We divide our experiments\footnote{For details about the experimental set up for evaluation, please refer to the supplementary material.} into two parts: 
\begin{enumerate}
    \item \textbf{Embedding Probing Tasks:}  The first part examines the effectiveness of the embeddings in encoding information such as genre, length, song-order etc.
    
    \item \textbf{Recommendation Task:} The second part evaluates the quality of embeddings for the purpose of recommendation.
\end{enumerate}{}

\subsection{Embedding Probing Tasks}

\subsubsection{Genre-Related Tasks}\label{sec:grtasks}

\begin{itemize}
    \item \textbf{Genre Prediction Task (GPred-Task)}\\
    This task measures to what extent a playlist embedding encodes the genre-related information of the songs it contains. Given a playlist embedding, the goal of the classifier is to predict the correct genre for the playlist. The task is formulated as multi-class classification, with nine output classes. Training samples are weighted-by-class as the dataset is skewed with the majority class (electronic) having 18,138 samples while the minority class (classical) just having 75 samples. \\
    
    \item \textbf{Genre Diversity Prediction Task (GDPred-Task)} \\
    This task measures the extent to which the playlist embedding captures the sense of the homogeneity/diversity of the songs (with regards to their genre) constituting it. Given a playlist embedding, the goal of the classifier is to predict the number of genres spanned by the songs in that playlist. The task is also formulated as multi-class classification, with 3 output classes being low diversity (0-3 genres), medium diversity (3-6 genres) and high diversity (6-9 genres).
\end{itemize}{}

\subsubsection{Playlist Length Prediction Task (PLen-Task)}\label{sec: ltask}
This task measures to what extent the playlist embedding encodes its length. Given a playlist embedding, the goal of the classifier is to predict the length (number of songs) in the original playlist. Following \cite{adi2016fine}, the task is formulated as multi-class classification, with ten output classes (spanning the range [30-250]) corresponding to equal binned lengths of size $20$. Training samples are class-weighted as the dataset is unbalanced, with a majority class (lengths 30-50 songs) having 78,015 samples and the minority class (230-250 songs) having just 1098 samples.

\subsubsection{Song-content Task (SC-Task)}
The song content \textbf{(SC)} closely follows the Word Content (WC) task described by \cite{conneau2018you} in testing whether it is possible to recover information about the original words in the sentence from its embedding. We pick 750 mid-frequency songs (the middle 750 songs in our corpus of songs sorted by their occurrence count ), and sample equal numbers of playlists that contain one and only one of these songs. We formulate it as a 750-way classification problem where the aim of the classifier is to predict which of the 750 songs does a playlist contain, given the playlist embedding.

\subsubsection{Song-order Related Tasks}

\begin{itemize}
    
    \item \textbf{Bigram-Shift Task (BShift-Task)}\\
    Text sentences are governed by language grammatical rules. These rules govern the existence of certain bi-grams in the language (e.g. \textit{will do}, \textit{will go} ) as well as the lack of existence of others (eg. \textit{try  will}). In the field of natural language processing, the Bigram Shift \textbf{(BShift)} experiment, introduced in \cite{adi2016fine}, is a very good way to measure the extent to which word-order information is encoded in the sentence embeddings.
    This evaluation task is formulated as a binary classification problem, where the aim of the classifier is to distinguish between original sentences from the corpus and sentences where two adjacent words have been inverted

    \item \textbf{Permute Classification Task} \\
Through this task we aim to answer the questions: Can the proposed embedding models capture song order, and if they can, to what extent? We split this task into two sub tasks: \textbf{i) Shuffle Task}, and  \textbf{ ii) Reversal} task. In the \textbf{Shuffle} task, for each playlist in our task-specific dataset\footnote{For details for the dataset as well as the experiment, please refer the supplementary material}, we randomly shuffle a fraction of all the songs in that playlist. We use two ways to shuffle the playlists: a) by selecting a random block in the playlist and shuffling just that block (\textbf{shuffle type-1}), and b) Randomly selecting songs from the playlist and shuffling them (\textbf{shuffle type-2}).  We then train a binary classifier where the aim of the classifier is to distinguish between an original and a permuted playlist. The \textbf{Reversal} task is similar to the Shuffle task except that the randomly selected sub-sequence of songs are reversed instead of shuffled. We further extend both these tasks with a slight modification that from the original dataset, playlists which are inverted are not included in the dataset and vice versa. 
\end{itemize}{}

\subsection{Recommendation Task}
This task is set up to evaluate our proposed approach for recommendation purposes by measuring the extent to which the playlist space created by the embedding models is relevant, in terms of the similarity of genre and length information of closely-lying playlists.\\
We populate an embedding space with the playlist embeddings using the Approximate Nearest Neighbors Algorithm, and Spotify ANNOY library\cite{bernhardsson2013annoy}. A playlist is randomly selected in the space and the returned search results are compared with the queried playlist in terms of genre and length information. There are nine possible genre labels, similar to the GPred-Task. For comparing length, classes are created as described in section \ref{sec: ltask}. An average of 100 precision values for each query is considered. High precision would signify good performance of the embeddings in terms of capturing information relevant to playlist recommendation. 

\section{Results}

\begin{table}
     \caption{Evaluation (Embedding Probing) task accuracies for the embedding models. The best result for each task is displayed in \textbf{bold font}.} 
    \centering
     \begin{tabular}{|| c | c | c | c | c ||}
          \hline
         \multicolumn{5}{||c||}{\textbf{Evaluation Tasks}} \\
         \hline
         \textbf{} & \textbf{GPred-Task} & \textbf{GDPred-Task} &  \textbf{PLen-Task}  & \textbf{SC-Task} \\ [0.5ex] 
          \hline
          \textbf{base-BoW Model} & 96.8 & 79 & 34 & 39.6 \\ 
           \hline
           \textbf{weighted-BoW Model} & \textbf{97.5} & \textbf{80.05} & 33.4 & \textbf{44.3} \\ 
           \hline
          \textbf{base-Seq2seq Model} & 76.6 & 75.8 & 70.7 & 15.3\\
         \hline
          \textbf{bi-Seq2seq Model} & 84.9 & 76.2 & \textbf{71.9} & 21.7\\
         \hline
        \end{tabular}
   
    \label{tab:results_table}
\end{table}

The results for the embedding probing tasks are presented in the Table \ref{tab:results_table}. In this section we provide a detailed description of our experimental results along with their analysis and insights. For each of the discussed tests – genre, length, and content – we investigate the performance of different embedding models across multiple embedding lengths. Results are shown in Figure \ref{fig:accuracy-emb-size}.

\subsection{Embedding Probing Task Results}
\subsubsection{Genre Task Results}\label{sec:grtask-results}
For the \textbf{GPred-Task}, BoW models outperform the Seq2seq models. This can be attributed to the reasoning that the playlist vector created by averaging the constituent songs is embedded in the space of the songs as their centroid. Since the genre of the playlist is the genre of its songs, the BoW models outperform the seq2seq at genre prediction. For the Seq2seq models, the performance appears to improve with increasing RNN hidden state size in the encoder, as larger embedding sizes generally have more space to encode sequence information. In the \textbf{GDPred-Task} as well, BoW-based models perform better than the Seq2seq models with the weighted-BoW model achieving 80\% accuracy while Seq2seq models achieve an accuracy of 76\%.

\subsubsection{Length Task Results}\label{sec:lrtask-results}
For the \textbf{PLen-Task}, the Seq2seq models perform quite well, achieving close to 72\% accuracy while the BoW models perform quite poorly managing just around 35\% accuracy. The performance of the Seq2seq models doesn't come as any surprise as it has been widely studied that seq2seq models are able to capture such characteristics of the sentences. Poor performance of the BoW models however is indeed surprising as BoW models have been shown to perform comparatively better on this task \cite{adi2016fine} \cite{conneau2018you}.

\subsubsection{Song Content Task Results}
For the \textbf{SC-Task}, the seq2seq models performed poorly compared to the BoW models. However, our results closely match the results for the WC Task for the unsupervised models in \cite{conneau2018you}. The authors cite the inability of the model to capture the content-based information due to the complexity of the way the information is encoded by the model.

\begin{figure}[!htb]
 \centerline{\framebox{
 \includegraphics[scale=0.3]{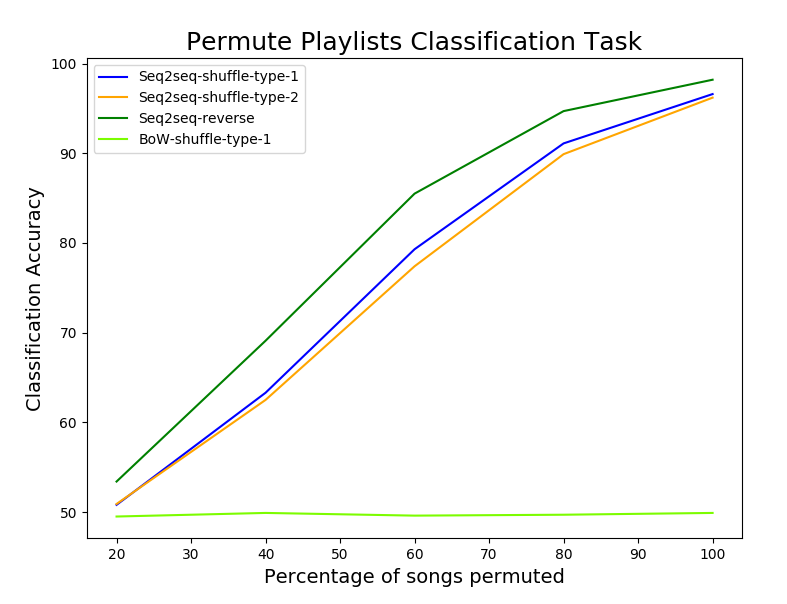}}}
 \caption{Classification accuracy vs Permute proportion for the Permute-classification task}
 \label{fig:shuffle}
\end{figure}

\subsubsection{Song Order Tasks Results}
For the \textbf{BShift-Task}, we get an accuracy of ~49\% for all the embedding models (both BoW and seq2seq), meaning the classifier is unable to distinguish the original playlist from bigram-inverted ones.
For the \textbf{Permute Classification Task}, as seen in Figure \ref{fig:shuffle}, the base Seq2seq model is able to distinguish correctly the permuted playlists from the original playlists as the proportion of the permutation is increased. Even for our extension of the tasks where complementary playlist pairs are not added to the dataset, the classifier can still distinguish between the original and the permuted playlists. On the other hand, BoW models cannot distinguish between the original and permuted playlists, making seq2seq models better for capturing the song-order in the playlist.\\

\subsection{Recommendation Task Results}
\begin{figure}[!htb]
 \centerline{\framebox{
 \includegraphics[scale=0.35]{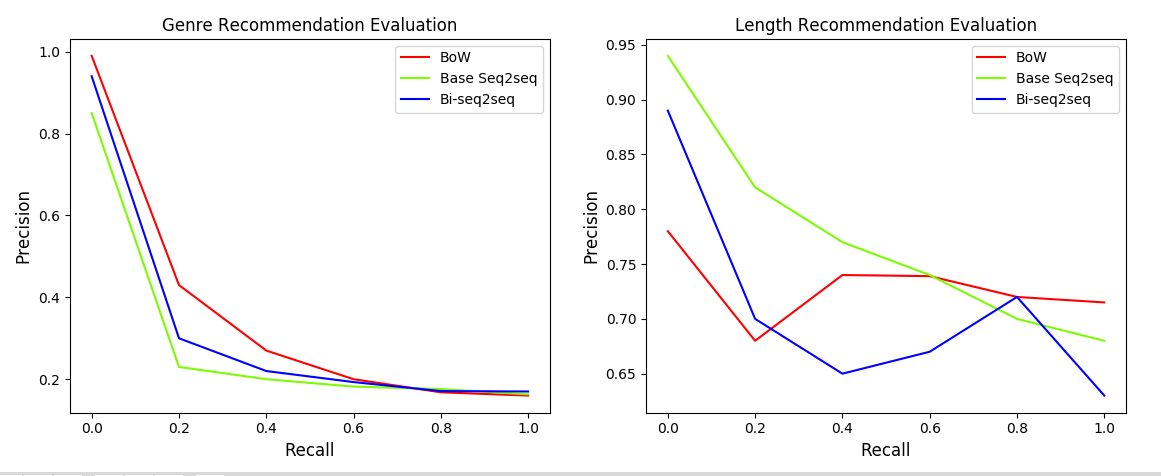}}}
 \caption{Precision-recall graph plotted for Genre and Length Recommendation task}
 \label{fig:recommend}
\end{figure}

The Recommendation task, as shown in Fig. \ref{fig:recommend}, captures some interesting insights about the effectiveness of different models for capturing different characteristics:
\begin{itemize}
    \item High precision values demonstrate the relevance of the playlist embedding space in terms of playlist-genre and length similarity of closely-lying playlists, which is the first and foremost expectation from a recommendation system.
    
    \item BoW models capture genre information\footnote{Since BoW created playlist embeddings lie in the song space (as calculated using arithmetic mean of song embeddings) where genre annotation happens, they perform better.} better than seq2seq models, confirming our genre-related findings from Section \ref{sec:grtask-results}.
    
    \item Length information is captured better by the Seq2seq models, confirming our length-related findings from Section \ref{sec:lrtask-results}
    
    \item The skewness of the graphs (especially for length-recommendation) can be attributed to the imbalance in the dataset.
\end{itemize}{}

\begin{table}[!htb]
     \caption{Encoder models performance comparison table. \checkmark \checkmark indicates better performance of the model compared to the other model. X indicates inability of the model to perform the task.} 
    \centering
     \begin{tabular}{|| c | c | c ||}
         \hline
         \textbf{}  & \textbf{BoW} & \textbf{Seq2seq} \\ [0.5ex] 
          \hline
          \textbf{Genre Information} & \checkmark\checkmark &  \checkmark \\ 
           \hline
           \textbf{Song Information} & \checkmark\checkmark &  \checkmark \\ 
           \hline
          \textbf{Length Information} & \checkmark &  \checkmark\checkmark \\
         \hline
          \textbf{Order Information} & X & \checkmark\checkmark \\
         \hline
          \textbf{Compute Complexity Required} & \checkmark\checkmark & \checkmark \\
         \hline
        \end{tabular}
    \label{tab:per_comparison}
\end{table}

\section{Results: Making Sense of Everything}
As shown in Table \ref{tab:per_comparison}, the evaluation task results capture some interesting details about the nature and capabilities of the models we use for this work to create playlist embeddings. BoW models perform quite well for genre and content-related tasks whereas seq2seq models capture the length and song-order information better. Going by these observations, we can use an ensemble of BoW and seq2seq models for building a playlist recommendation engine, having the BoW model create an initial subset of recommendable playlists based on genre information from the corpus, and seq2seq models refine that list to find the most similar playlists in terms of song-order and length information. \\
Another important point to consider is that the seq2seq model would not work for playlists having all the songs from outside the corpus which the model is trained on. The reason is that the current model derives the semantic characteristics of the songs and the playlists based solely on the co-occurrence of the songs in the corpus. To mitigate this problem, additional information needs to injected into the system such as audio content information, lyric-based information, or a combination of both, such that it can be used by the encoder to learn semantic characteristics of songs independent of their neighborhood in which they occur.

\section{Conclusion}
We have presented a seq2seq based approach for learning playlist embeddings, which can be used for tasks such as playlist discovery and recommendation. First, we define the problem of learning a playlist-embedding and describe how we formulate it as a seq2seq-based problem. We compare our proposed model with two BoW-based models on a number of tasks chosen from the field of natural language processing (like sentence length prediction, bi-gram shift experiment etc.) as well as music (genre prediction and genre-diversity prediction). We show that our proposed approach is effective in capturing the semantic properties of playlists. We also evaluate our approach using a recommendation task, showing the effectiveness of the learned embeddings for recommendation purposes. Our approach can also be extended for learning even better playlist-representations by integrating content-based (lyrics, audio etc.) song-embedding models, and for generating new playlists by using variational sequence models.

\bibliographystyle{unsrt}  
\bibliography{references}  






\clearpage
\begin{center}
\textbf{\large Supplemental Material: Representation, Exploration and Recommendation of Music Playlists}
\end{center}
\setcounter{equation}{0}
\setcounter{figure}{0}
\setcounter{table}{0}
\setcounter{page}{1}
\setcounter{section}{0}
\section{Data Statistics}\label{data-stats}
\begin{figure}[!htb]
 \centerline{\framebox{
 \includegraphics[scale=0.5]{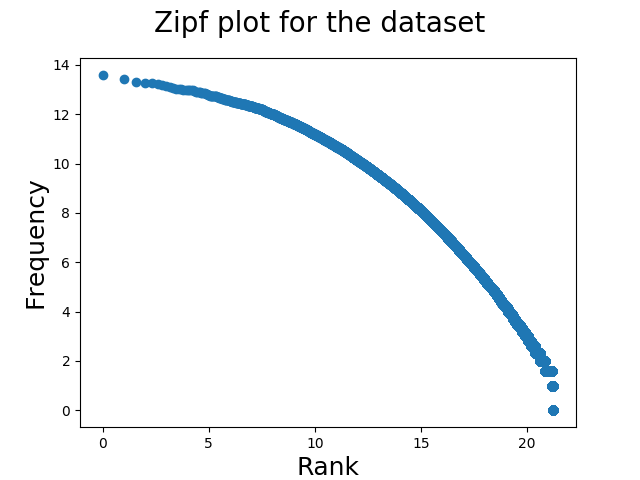}}}
 \caption{Zipf plot for the corpus. Log scale used for frequency and rank}
 \label{fig:zipf}
\end{figure}

The data distribution roughly follows the Zipf's law. We plotted a Zipf plot of our corpus as shown in Figure \ref{fig:zipf}.

\subsection{Playlists Length Statistics}
More than half of the playlists (401,880) are of length less than 50 and another 246,427 (33\%) of the playlists have lengths in the range [50-150]. Statistics related to the length of the playlists are given in Table \ref{tab:len_stats_table}.
\begin{table*}[!htb]
    \caption{Corpus Length Statistics}
    \centering
     \begin{tabular}{|| c | c | c | c | c ||}
          \hline
         \multicolumn{5}{||c||}{\textbf{Playlist Corpus Length statistics}} \\
         \hline
         \textbf{Mean} & \textbf{Std} & \textbf{Median} & \textbf{Min.} & \textbf{Max.}\\[0.5ex] 
          \hline
           83.3 & 133 & 45 & 10 & 5000\\
         \hline
        \end{tabular}
    \label{tab:len_stats_table}
\end{table*}

\subsection{Genre Homogeneity/Diversity Statistics}

Out of 745,543 playlists, 49,164 playlists have all of their songs genre annotated. Out of 49,164 playlists, 24,422 (~49\%) playlists have less than or equal to 3 genres in total across all of their songs, 23,162 (~47\%) playlists have less than of equal to 6 genres in total, and 1580 (~3\%) playlists have more than 6 genres. 

\section{Word-2-vec: Set up and Training}\label{word2vec}
We train a word2vec model on our corpus to get the song embeddings, using the Gensim\cite{rehurek_lrec} implementation. We use the Skipgram algorithm with negative sampling value set to 5 and window size of 5 as well. Minimum threshold for the occurrence of words is set to 5. We train the word vectors of sizes k $\in$ \{500, 750, 1000\}.

\section{Experimental Setup}
\subsection{Training}
We used OpenNMT-lua library \cite{2017opennmt} as the Neural Machine Translation implementation library and AWS p3.16X large for conducting our experiments. Only the playlists with length less than 50 are are considered. Also, words with occurrence count less than 20 are discarded. This brings the vocabulary size down to 82,259 words and the number of playlists in the training set to 377,362. The training is done for 15 epochs and perplexity is used as the evaluation metric for the model. Batch-size was varied in the range [32-128] depending on the hidden state size and the memory constraints.

\subsubsection{Song Order experiments}
\begin{itemize}
    \item \textbf{Bigram shift: Data}\\
    To create the dataset for this experiment, we selected a list of 55265 playlists whose length lie in the range [50-100]. For each of these playlists, we created an additional playlist with two adjacent songs inverted. This resulted in a balanced dataset of size 110530.
    
    \item \textbf{Permute Shuffle}
    To create the dataset for this experiment, we selected a list of 38168 playlists whose lengths lie in the range [50-100].
    
\end{itemize}{}

\subsection{Evaluation}
A 1-hidden layer neural network implemented in Keras\cite{chollet2015keras} is used for all experiments. Each network is tailored for each task. Sklearn\'s \cite{scikit-learn} class weight computation library is used to learn class weights. We use categorical Cross Entropy loss with Softmax non linearity for all multi-class classification tasks and Binary Cross Entropy loss and Sigmoid non linearity for the binary classification tasks. 

\section{Other Tried Experiments}
\subsection{Genre Prediction (Multi-Label Prediction)}
This task \textbf{(GMLPred-Task)}  is an extension to the \textbf{GPred-Task} and the aim for this task is to predict all the genres given a playlist embedding. The task is formulated as multi-label classification, with the same nine output classes as used in the GPred task. Results in Table \ref{tab:gmlpred}.
 
\begin{table*}
     \caption{Evaluation accuracies for Multi-Label Genre Prediction task.}
    \centering
     \begin{tabular}{|| c | c | c | c ||}
          \hline
         \multicolumn{4}{||c||}{\textbf{Evaluation results for Genre Prediction Task}} \\
         \hline
         \textbf{base-BoW} & \textbf{weighted-BoW} & \textbf{base seq2seq} & \textbf{bi-seq2seq}\\[0.5ex] 
          \hline
           82.5 & 84 & 80.8 & 82.3\\
         \hline
        \end{tabular}
    \label{tab:gmlpred}
\end{table*} 
 
\subsubsection{Paraphrase detection Evaluation Task}
We attempted the Paraphrase detection Evaluation task \cite{dolan2004unsupervised} by creating a dataset of playlists sampled from the dataset and as their paired playlists, we chose:
\begin{enumerate}
\item A shuffled portion of the songs of the original playlist as the \textit{entailment} playlist.
\item A completely different playlist  with non overlapping songs as the \textit{contradictory playlist}
\end{enumerate}

We found that the model was very easily able to correctly tag the paired playlists. We realized that in our case, since the entailed playlists are always going to be shorter in length than the original playlists, the model can just focus on that for making the prediction while completely ignoring the content of the playlists. This experiment especially points towards the lack of supervised datasets in music.

\subsubsection{Genre Switch Prediction}
A genre switch is defined here as change in genre going from one song to another. A homogeneous playlist would have fewer number of such genre shifts than a more diverse playlist. The aim of this task \textbf{(GSPred-Task)} is to predict the number of all the genres switches given a playlist embedding. For this task, the absolute number of switches for a playlist is normalized by dividing it by the length of the playlist such that the final label lies between 0 and 1. The task is then formulated as multi-class classification, with the five output classes being low switch playlist (0-0.34), mid-switch playlist (0.34-0.67) and high-switch playlist (0.76-1.00). Results in Table \ref{tab:gspred}

\begin{table*}[!htb]
     \caption{Evaluation accuracies for Genre-shift Prediction task.}
    \centering
     \begin{tabular}{|| c | c | c | c ||}
          \hline
         \multicolumn{4}{||c||}{\textbf{Evaluation results for Genre-shift Prediction task}} \\
         \hline
         \textbf{base-BoW} & \textbf{weighted-BoW} & \textbf{base seq2seq} & \textbf{bi-seq2seq}\\[0.5ex] 
          \hline
           76.7 & 77.9 & 58.9 & 56\\
         \hline
        \end{tabular}
   \label{tab:gspred}
\end{table*}

\subsection{LSTM vs GRU comparison}
For training our models, we experimented with LSTM and GRU units. For almost all the experiments across all models and embedding sizes, LSTM unit performed better than GRU unit except for the Permute task where the GRU model(s) outperformed the LSTM-based models for all the embedding sizes.

\end{document}